\documentclass[aps,preprintnumbers,amsmath,amssymb,nofootinbib,preprint]{revtex4}
\usepackage[latin9]{inputenc}
\usepackage{hyperref}
\hypersetup{
    colorlinks,
    citecolor=blue,
    filecolor=blue,
    linkcolor=blue,
    urlcolor=blue
}
\date{\today}
\usepackage{physics}
\usepackage{amsmath,amsthm,amssymb}
\usepackage{xcolor}
\usepackage{graphicx}
\renewcommand{\v}[1]{ \ensuremath{ {\underline{#1}} }}

\begin{document}
\title{  Gluon Quasi Particles and the CGC Density Matrix    }

\author{Haowu Duan}
\affiliation{North Carolina State University, Raleigh, NC 27695, USA}
\author{Alex Kovner}
\affiliation{Physics Department, University of Connecticut, 2152 Hillside Road, Storrs, CT 06269, USA}
\author{Vladimir V. Skokov}
\email{VSkokov@ncsu.edu}
\affiliation{North Carolina State University, Raleigh, NC 27695, USA}
\affiliation{RIKEN-BNL Research Center, Brookhaven National Laboratory, Upton, NY 11973, USA}
\begin{abstract}
We revisit and extend the calculation of the density matrix and entanglement entropy of a Color Glass Condensate by including the leading saturation corrections in the calculation. We show that the density matrix is diagonal in the quasi particle basis, where it has the Boltzmann form. The quasi particles in a wide interval of momenta behave as massless two-dimensional bosons with the temperature proportional to the typical semi-hard scale $T=Q_s/\sqrt{\alpha_sN_c}$. Thus the semi-hard momentum region $Q_s<k<Q_s/\sqrt{\alpha_sN_c}$ arises as a well-defined intermediate regime between the perturbatively hard momenta and the nonperturbative soft momenta $k<Q_s$ in the CGC description of a hadronic wave function.
\end{abstract}

\maketitle
\tableofcontents
\newpage
\section{Introduction}

Recently there has been an increased interest in incorporating the concepts and methods of quantum information theory into nuclear and particle physics \cite{Klco:2021lap}. In particular  various aspects of entanglement in application to  
hadronic collisions have been considered in  \cite{Robin:2020aeh,Beane:2019loz,Armesto:2019mna,Neill:2018uqw,Kovner:2018rbf,Hagiwara:2017uaz,Hatta:2015ggc}.
In the context of high energy collisions various ideas about possible relevance of entanglement to thermalization and parton distributions have been discussed in \cite{Berges:2017hne,Berges:2017zws,Kharzeev:2017qzs,Gotsman:2020bjc}. The entanglement entropy between strongly coupled nonpetrubative modes and partonic components of hadronic wave function was conjectured to be the origin of the Boltzmann entropy of particles produced in the collisions.

It was pointed out in \cite{Kovner:2015hga} that the Color Glass Condensate (CGC) effective theory provides an explicit and calculable model of entanglement in high energy hadronic wave function.
The concept of entanglement implies partition of a system into the system of interest and its compliment.
In \cite{Kovner:2015hga}  the soft gluon degrees of freedom were considered as the system of interest  while the valence degrees of freedom (larger x partons) in the hadron wave function were treated as the complement. This partition is directly relevant to measurements  of observables at mid rapidity  which reflect the properties of the soft gluons.
Integrating
out the valence degrees of freedom in the original (pure state) hadron wave function
produces a mixed state density matrix of the soft gluons. The entropy associated with this mixed state is
the entanglement entropy in question.

Originally, the entanglement entropy between valence
degrees of freedom and soft gluons was calculated in \cite{Kovner:2015hga} in  the dilute limit, $Q_s^2/k^2 \ll 1$.
In the present paper we extend this calculation by accounting for saturation effects in a  ``mean field'' approximation.
In addition, we discuss some properties of the CGC reduced density matrix and the associated entropy which were not investigated in earlier work. In particular, we notice that the entropy has a Boltzmann form. This implies that the associated reduced density
matrix  not only can be diagonalized (this, of course, can always be done in principle), but also that the
eigenvalues of the density matrix are given by the powers of the same single number (which in our
case is a function of transverse momentum). In other words in the appropriate basis the reduced density matrix has a Boltzmann form albeit with a momentum-dependent effective temperature. We explicitly find this basis by performing a
Bogoliubov transformation from the original free gluon basis and discuss some interesting properties of the corresponding quasi-particles.

We start in Sec. II by reviewing the derivation of \cite{Kovner:2015hga} while also providing some technical details. In Sec. III we discuss the extension of this calculation by including the saturation effects. In Sec. IV we diagonalize the density matrix and give explicitly its Boltzmann form in the quasi particle basis. We conclude with a short discussion in Sec. V.

\section{Review: the entanglement in the Color Glass Condensate}

\subsection{The CGC hadron wave function}
For an ultrarelativistic proton, a large fraction of the longitudinal momentum is
carried by the valence degrees of freedom. When boosted, the valence partons
radiate gluons with lower longitudinal momentum which have a relatively short lifetime.
It is then natural to separate the degrees of freedom according to their longitudinal momentum: the large longitudinal momentum partons can be treated as static sources of soft, low longitudinal momentum, gluons.
These degrees of freedom are of course strongly correlated, and these correlations play an important role for the phenomenology. For example triggering on high multiplicities at forward rapidities also selects events with high multiplicity at mid rapidity.
A less conventional quantity which measures the correlation strength is the entanglement entropy -- the main focus of this paper.

In the CGC approach the hadron state vector can be written in the form
\begin{equation}
| \psi \rangle = | s \rangle  \otimes   | v \rangle
\end{equation}
where $| v \rangle $ is the  state vector characterizing the valence degrees of freedom and
$| s \rangle$ -- the state in the soft gluon Hilbert space. The direct product in this equation is not mathematically precise, as we alluded to
before, since the soft gluons are sourced by valence degrees of freedom. In CGC, this is encoded by
\begin{equation}
	\label{Eq:SoftVacuum}
    | s \rangle=\mathcal{C}| 0\rangle
\end{equation}
with the coherent operator
\begin{equation}
    {\cal C}=\exp\left\{2i\int_{\v{k}}\tr  b^i(\v{k})[a_i^+(\v{k})+a_i(-\v{k})]\right\}\,,
\end{equation}
where the summation over all color is implied.  Here $|0\rangle$ is the Fock vacuum of the soft gluon Hilbert space. The ``background'' Weizs\"acker-Williams gluon field $b^i_a$ is the solution of the static Yang-Mills equation
\begin{equation}
	\partial_i b^{a}_i(x) = g\rho^a(x)\,.
\end{equation}
where $\rho^a(x)$ is the color charge density of the valence gluons.

Eq.~\eqref{Eq:SoftVacuum} strictly speaking is valid in the regime when  the
color charge density is weak where one can perform the  perturbative diagonalization of the
QCD Hamiltonian in the soft gluon sector. Nevertheless in principle the solution to Eq.~\eqref{Eq:SoftVacuum} contains nonlinearities in $\rho$ which reflect certain gluon saturation effects.
Often, for the sake of simplification,  one considers the limit  $k\gg Q_s$ where these nonlinearities are small. In this case,  the solution of the
Yang-Mills equation can be expressed as\footnote{For completeness we provide the conventions for the Fourier transformation in two dimensions:
\begin{align}
    \notag
    f(\v{x})&=\int \frac{d^2\v{k}}{(2\pi)^2}  e^{i\v{k}\cdot \v{x}}f(\v{k})  \equiv \int_\v{k}  e^{iv{k}\cdot \v{x}} f(\v{k}) \\
    \notag
    f(\v{k})&=\int d^2x  e^{-i\v{k} \cdot \v{x}}f(\v{x})\,.
\end{align}In this paper $\v{k}$ stands for 2-d vector and k stands for its magnitude.}
\begin{equation}\label{Eq:wwfield}
    b^i_a(\v{k})=g\rho_a(\v{k})\frac{i\v{k}_i}{k^2} + c_a^i(\v{k})\,,
\end{equation}
where $c_a^i(k)$ is at least ${\cal O}(\rho^2)$. In this section, following the original derivation of Ref.~\cite{Kovner:2015hga},
we neglect the contribution due to $c_a^i(k)$. We will come back to consider nonlinear terms  in the next section.  Note that the first term in Eq.~\eqref{Eq:wwfield} is longitudinal (in the two dimensional sense) while $c^i$ is transverse, that is $c^i k^i = 0$. Thus neglecting $c_a^i$
leads to excitation of only longitudinal gluon degrees of freedom in the soft gluon wave function.

We use the McLerran-Venugopalan model to model the valence state \cite{McLerran:1993ni,McLerran:1993ka}. This corresponds to treating the color charges as the only relevant valence degrees of freedom with the distribution
\begin{equation}
  \langle \rho | v \rangle \langle v |  \rho \rangle=   e^{-\int_{\v{k}}
\rho_a(\v{k})
  \frac{1}{2\mu^2(\v{k})}\rho^*_a(\v{k})}\,.
\end{equation}

The density matrix of the complete system is thus given by
\begin{equation}
    \hat{\rho}=|s\rangle \langle s| \otimes |v\rangle \langle v| \,,
\end{equation}
and corresponds to a pure state, as it should.

\subsection{The reduced density matrix of soft gluons}
To proceed with evaluation of the entanglement entropy, we integrate out the valence modes. The resulting  reduced density matrix for the soft modes is defined as
\begin{align}
    \hat{\rho}_r=
    \sum_v \langle v| \hat \rho |v\rangle  =
    \mathcal{N}\int D \rho\, \,  e^{-\int_{\v{k}}\frac{1}{2\mu^2(\v{k})}\rho_a(\v{k})\rho^*_a(\v{k})}\mathcal{C}|0\rangle \langle0|\mathcal{C}^\dagger \, .
\end{align}
Our next step is to compute the reduced density matrix  explicitly.
For this it is convenient to
introduce the notation
\begin{align}
 \notag
 \phi(\v{k})&=a(\v{k})+a^+(-\v{k})\,, \\
 \notag
 \phi(x)&=a(x)+a^+(x)\,.
\end{align}

The {\it matrix element} of the reduced density matrix in the field basis is
\begin{align}\label{matel}
   \langle \phi_1 | \hat{\rho}_r| \phi_2 \rangle =
   \mathcal{N}\int D \rho\, \,  e^{-\int_{\v{k}}\frac{1}{2\mu^2(\v{k})}\rho_a(\v{k})\rho^*_a(\v{k})}
   \langle \phi_1 |\mathcal{C}(\rho,\phi)|0\rangle \langle0|\mathcal{C}^\dagger(\rho,\phi) | \phi_2 \rangle\,.
\end{align}
In the above
\begin{align}
 \langle \phi_1 |\mathcal{C}(\rho,\phi)|0\rangle  =
 \langle \phi_1 |  \exp\left\{i\int_k b_a^i(\v{k}) \phi_a^* (\v{k}) \right\}
 |0\rangle
  = \exp\left\{i\int_k b_a^i(\v{k}) \phi_{1 a}^* (\v{k}) \right\}   \langle \phi_1 |0\rangle\,.
 \end{align}
 The wave function for the coherent vacuum  $\langle \phi_1 |0\rangle$ is known
 and we will use its explicit form at a later stage of the derivation.
 Eq.~\eqref{matel} then becomes
 \begin{align}
   \langle \phi_1 | \hat{\rho}| \phi_2 \rangle =
   \mathcal{N}\int D \rho\, \,  e^{-\int_{\v{k}}\frac{1}{2\mu^2(\v{k})}\rho_c(\v{k})\rho^*_c(\v{k})}
   e^{i\int_{\v{k}} b_a^i(\v{k}) \phi^{*i}_{1,a} (\v{k})  }
   \langle \phi_1 |0\rangle \langle0|\phi_2 \rangle
   e^{-i\int_{\v{k}} b_{b}^{*j}(\v{k}) \phi^{j}_{2,b} (\v{k}) }
   \,.
\end{align}
The integral over the color charge density $\rho$ is Gaussian and can be evaluated in a straightforward manner.  Using Eq.~\eqref{Eq:wwfield} and neglecting nonlinear terms in the solution we obtain
 \begin{align}
     \label{Eq:Gaussian}
     \notag
    &  \mathcal{N} \int D \rho\, \,  e^{-\int_{\v{k}}\frac{1}{2\mu^2(\v{k})}\rho_c(\v{k})\rho^*_c(\v{k}) + i\int_{\v{k}} b_a^i(\v{k}) \phi^{*i}_{1,a} (\v{k})  -i\int_{\v{k}} b_{b}^{*j}(\v{k}) \phi^{j}_{2,b} (\v{k}) }
    \\ &=
          \mathcal{N} \int D \rho\, \,  e^{-\int_{\v{k}}\frac{1}{2\mu^2(\v{k})}\rho_c(\v{k})\rho^*_c(\v{k}) + i\int_{\v{k}} b_a^i(\v{k}) (\phi^{i}_{1,a} (- \v{k}) - \phi^{i}_{2,a} (- \v{k}))   }
 \notag \\
 \notag
 &=
  \mathcal{N}
  \int D \rho\, \,  e^{-\int_{\v{k}}
   \frac{1}{2\mu^2(\v{k})}
  \left[ \rho_a (\v{k})
- g\mu^2  \frac{\v{k}_i}{k^2}
(\phi_{a1i}(\v{k}) - \phi_{a2i}(\v{k}))
\right]
\left[ \rho_a (-\v{k})
- g\mu^2  \frac{- \v{k}_i}{k^2}
 (\phi_{a1i}(-\v{k}) - \phi_{a2i}(-\v{k}))
 \right]
}
\\
 \notag
 & \times
 e^{-\int_{\v{k}}
\frac{g^2\mu^2}{2} \frac{\v{k}_i \v{k}_j}{k^4}   (\phi_{a1j}(\v{k}) - \phi_{a2j}(\v{k})) (\phi_{a1i}(-\v{k}) - \phi_{a2i}(-\v{k}))
 }
   \\
    &=
    e^{-\int_{\v{k}}
   \frac{g^2\mu^2}{2} \frac{\v{k}_i \v{k}_j}{k^4}   (\phi_{a1j}(\v{k}) - \phi_{a2j}(\v{k})) (\phi_{a1i}(-\v{k}) - \phi_{a2i}(-\v{k}))
    }\,.
    \end{align}
Therefore the matrix element reads
 \begin{align}
     \label{Eq:rhoMEprel}
   \langle \phi_1 | \hat{\rho}| \phi_2 \rangle
   & =
 \langle \phi_1 |0\rangle  \langle0|\phi_2 \rangle
 e^{-\int_{\v{k}}
\frac12 M^{ab}_{ij}(\v{k})  (\phi^a_{1j}(\v{k}) - \phi^a_{2j}(\v{k})) (\phi^b_{1i}(-\v{k}) - \phi^b_{2i}(-\v{k}))
 }\,
\end{align}
with
\begin{equation}
 M^{ab}_{ij}(\v{k}) \equiv g^2 \mu^2(k)\delta^{ab}\frac{\v{k}_i\v{k}_j}{k^4}
 \end{equation}
This has to be supplemented by the vacuum wave function (in terms of fields in the coordinate space representation):
\begin{equation}
\langle \phi |0\rangle = N_{\text vac} e^{- \frac 14  \int_x\phi_i^a(x) \phi_i^a(x) },
\end{equation}
where $N$ is defined by the condition:\footnote{The argument of the exponential of the vacuum wave function is normalized to yield
the same result for the density matrix as we previously obtained in  \cite{Kovner:2015hga}, see also \cite{Duan:2020jkz}
and \cite{Duan:2021kmf}.}
$$
1 = \langle 0| 0 \rangle  = \int D \phi \langle 0| \phi \rangle  \langle \phi | 0 \rangle
 = N_{\text vac}^2 \int  D \phi e^{- \frac12 \int_x \phi_i^a(x) \phi_i^a(x) } \,.
$$

We finally obtain that the matrix element has the follwoing form:
 \begin{align}
     \label{Eq:rhoME}
     \langle \phi_1 | \hat{\rho}| \phi_2 \rangle
     & =
     N^2_{\text vac}
   e^{-\int_{\v{k}}
  \left[
  \frac12 M^{ab}_{ij}(\v{k})  (\phi^a_{1j}(\v{k}) - \phi^a_{2j}(\v{k})) (\phi^b_{1i}(-\v{k}) - \phi^b_{2i}(-\v{k}))
  + \frac{1}{4} \phi^a_{1i}(\v{k}) \phi^a_{1i}(-\v{k}) + \frac{1}{4} \phi^a_{2i}(\v{k}) \phi^a_{2i}(-\v{k})
  \right]
   }\,.
\end{align}

The representation \eqref{Eq:rhoME} is the most convenient  for computing the von Neumann entropy.


\subsection{The von Neumann entropy}

The von Neumann entropy of a density matrix is defined as\footnote{For a thorough discussion of various definitions of entropy see \cite{IJpelaar:2021oyw}.} :
\begin{equation}
   S^E=-\tr[\hat{\rho}\, \ln \hat{\rho} ]
\end{equation}
The calculation is facilitated by using the replica trick.
Using the identity:
\begin{align}
    \ln \hat{\rho} = \lim_{\epsilon \rightarrow 0}\frac{1}{\epsilon}(\hat{\rho}^{\epsilon}-1)
\end{align}
we have:
\begin{align}
     S^E=-\lim_{\epsilon \rightarrow 0}
     \tr(\frac{\hat{\rho}^N-\hat{\rho}}{\epsilon})
     = -\lim_{\epsilon \rightarrow 0}
     \frac{\tr  \hat{\rho}^N  - 1 }{\epsilon}
     \, ,
\end{align}
where $N = \epsilon+1$. Thus the problem of evaluating the von Neumann entropy reduces to  computation of  $\tr \hat{\rho}^N$ for integer $N$ and subsequent analytic continuation to arbitrary $N$.
Note that $\tr \hat{\rho}^N$  is related to the $N$-th Renyi entropy
\begin{equation}
	S_N = \frac{1}{1-N}
	\ln \left[ {\rm Tr}   \hat\rho  ^N   \right]
    \,.
	\label{Eq:RenN}
 \end{equation}

It is straightforward to proceed:
$$
{\rm Tr} \,  \hat \rho  ^N
 = \int D \phi_{\bf 1} \langle \phi_{\bf 1}|  \hat \rho  ^N | \phi_{\bf 1}  \rangle
 = \int D \phi_{\bf 1} D \phi_{\bf 2}  \langle \phi_{\bf 1}| \hat \rho | \phi_{\bf 2} \rangle  \langle \phi_{\bf 2}| \hat \rho  ^{N-1} | \phi_{\bf 1}  \rangle = \ldots = \int \prod_{n=1}^N  D \phi_{\bf n}
 \langle \phi_{\bf n}| \hat \rho | \phi_{\bf n+1} \rangle \,,
$$
where the fields satisfy periodic boundary conditions in replica space $\phi_{\bf  N+1} = \phi_{\bf 1}$.
We use the boldface font to denote the field index.
With the help of Eq. \eqref{Eq:rhoME}, we can explicitly write
\begin{align}
&    \Tr \hat{\rho}^N =  N_{\text vac}^N \int  D\phi_{\bf 1} D\phi_{\bf 2} \ldots D\phi_{\bf N} \notag \\ & \times
    \exp \bigg\{-\frac{1}{2} \sum_{{\bf n}=1}^N \int_{\v{k}}\phi_{{\bf n}i}^{a} (\v{k})\phi_{{\bf n}i}^{a} (-\v{k})
\notag \\ &
    -\frac{1}{2}\sum_{{\bf n}=1}^N \int_{\v{k}}
    \left[ \phi_{{\bf n}i}^{a}(\v{k}) - \phi_{{\bf (n+1)}i}^{a} (\v{k}) \right] M^{ab}_{ij} (\v{k}) \left[ \phi_{{\bf n}j}^{b}(-\v{k})- \phi_{{\bf (n+1)}j}^{b}(-\v{k})  \right] \bigg\}\,.
\end{align}
The integrand involves mixing terms between different replica fields, however it can be diagonalized by performing the Fourier transformation with respect to the replica index (we suppress other indices for simplicity):
\begin{align}
    \tilde{\phi}_\eta &=\frac{1}{N}\sum_{{\bf n}=1}^N e^{i\frac{2\pi}{N} {\bf n}   \eta }\phi_{{\bf n}};
    \\
     \phi_{{\bf n}}&=\sum_{\eta=0}^{N-1} e^{-i\frac{2\pi}{N} {\bf n} \eta}\tilde{\phi}_\eta\,.
\end{align}
This yields
\begin{align}
     \sum_{{\bf n}=1}^N ( \phi_{{\bf n} i}^a- \phi_{{\bf(n+1)} i}^a)
     (\phi_{ {\bf n} j}^b- \phi_{{\bf(n+1)}j}^{b})&
     = N\sum_{\eta=0}^{N-1}(1-e^{-i\frac{2\pi \eta}{N}})(1-e^{i\frac{2\pi \eta}{N}}) \tilde{\phi}^a_{\eta i}\tilde{\phi}^b_{(-\eta)j}\\
     &=4N\sum_{\eta=0}^{N-1}\sin^2 \left(\frac{\pi}{N} \eta \right)\tilde{\phi}_{\eta i}^a \tilde{\phi}_{(-\eta) j}^b
\end{align}
and the problem is reduced to a standard Gaussian integral:
\begin{align}
    \Tr\, \hat{\rho}^N &= 
{\cal N}
    \int  D\tilde{\phi}_0
    D\tilde{\phi}_1 \ldots D\tilde{\phi}_{N-1}
    \notag \\ & \times \exp \bigg\{-N \sum_{\eta=0}^{N-1} \tilde{\phi}^a_{\eta,i}
    \left[  \frac12 \delta_{ij}\delta^{ab}   + 2M^{ab}_{ij}\sin^2\left(\frac{\pi}{N}n\right)  \right]\tilde{\phi}_{(-n),j}^b  \bigg\}\,,
\end{align}
where we have absorbed the Jacobian into the normalization factor which we will establish below.
The Gaussian integral  yields:
\begin{align}
     \Tr\, \hat{\rho}^N & = {\cal N} \det \left[\prod_{\eta=0}^{N-1}\left(\frac12+2M \sin^2\left(\frac{\pi}{N}\eta\right)\right)^{-\frac{1}{2}}  \right]\\
     &={\cal N} \det[\prod_{\eta=0}^{N-1} \left(\frac12+M\left(1-\cos\left(\frac{2\pi}{N}  \eta \right)\right)\right)^{-\frac{1}{2}}    ]\,.
     \label{Eq:det}
\end{align}

The matrix $M$ in Eq.~\eqref{Eq:det} is diagonal in color, momentum, and the replica space. Its polarization structure is purely longitudinal, so that the eigenvalues are
$M_-=0$ and $M_+=\frac{g^2 \mu^2}{k^2}$.
We therefore get
\begin{align}
     \Tr\, \hat{\rho}^N
     &= {\cal N} \det[\prod_{\eta=0}^{N-1} \left(\frac12+M_-\left(1-\cos\left(\frac{2\pi}{N}  \eta \right)\right)\right)^{-\frac{1}{2}}    ]^{N_a}
     \notag
     \\ &\times
     \det[\prod_{\eta=0}^{N-1} \left(\frac12+M_+\left(1-\cos\left(\frac{2\pi}{N}  \eta \right)\right)\right)^{-\frac{1}{2}}    ]^{N_a}
     \notag
     \\ &= {\cal N} \det[\prod_{\eta=0}^{N-1} \left(1+2M_+\left(1-\cos\left(\frac{2\pi}{N}  \eta \right)\right)\right)    ]^{-N_a/2}\,,
     \label{Eq:det2}
\end{align}
where we again absorbed irrelevant constants into ${\cal N}$.

To perform summation over $\eta$ we adapt the formula (1.394) from \cite{alan2007table}
\begin{equation}
    \prod_{l=0}^{N-1} \left[ x^2 - 2 xy \cos \frac{2 l \pi}N + y^2  \right]
    = (x^{N} - y^N)^2
\end{equation}
using the following mapping
\begin{equation}
    x^2 + y^2 =  1+2 M_+, \notag \quad
    2 x y = 2 M_+\,.
\end{equation}
The result is:
\begin{align}
    \prod_{\eta=0}^{N-1} \left(1+2M_+\left(1-\cos\left(\frac{2\pi}{N}  \eta \right)\right)\right) =
    \frac{1}{2^{2N}} \left[ \left(\sqrt{1+4 M_+} +1\right)^N - \left(\sqrt{1+4 M_+} -1\right)^N   \right]^2\,.
\end{align}
Thus we arrive at
\begin{align}
     \Tr\, \hat{\rho}^N
     &=
     {\cal N}
     \det _k \left\{ 2^{N N_a} \left[ \left(\sqrt{1+4 M_+} +1\right)^N - \left(\sqrt{1+4 M_+} -1\right)^N   \right]^{-N_a}
     \right\}
     \,.
     \label{Eq:det3}
\end{align}
where $\det_k$ denotes determinant in momentum space (product over momenta) only, and $N_a=N_c^2-1$.
The normalization factor ${\cal N}$ can now be determined by requiring that the reduced density matrix is properly normalized. Setting  $N=1$ we have  $\Tr \hat \rho= {\cal N} \det_k 1$ and thus ${\cal N}=1/\det_k 1$.

Performing analytical continuation  $N=1+\epsilon$ and expanding in $\epsilon$, we obtain:
\begin{align}
     \Tr\, \hat{\rho}^N
     & \approx
     {\cal N}
     \det \left\{
     1 - \epsilon \frac{N_a}{2}  \left( \ln M_+ + \sqrt{ 1 + 4 M_+ }
     \ln \left[ 1 +  \frac{1}{2 M_+} + \frac{1}{2 M_+} \sqrt{ 1 + 4 M_+ }  \right]   \right)
     \right\}
     \notag \\
     & \approx
     1 -\epsilon \frac{N_a}{2}  S_\perp \int_{\v{k}} \left( \ln M_+ + \sqrt{ 1 + 4 M_+ }
     \ln \left[ 1 +  \frac{1}{2M_+} + \frac{1}{2M_+} \sqrt{ 1 + 4 M_+ }  \right]   \right)\,.
     \label{Eq:det4}
\end{align}

And therefore the entanglement entropy is given by
\begin{align}\label{Eq:ent}
    S_E    = \frac{N_a}{2} S_\perp \int_{\v{k}} \left( \ln M_+ + \sqrt{ 1 + 4 M_+ }
    \ln \left[ 1 +  \frac{1}{2 M_+} + \frac{1}{2 M_+} \sqrt{ 1 + 4 M_+ }  \right]   \right)\,.
\end{align}
With the definition $M_+ = \frac{g^2 \mu^2}{k^2}$,  this reproduces the result of Ref.~\cite{Kovner:2015hga} \footnote{ \cite{Kovner:2015hga}  used a different normalization of $\mu$, see also Ref.~\cite{Duan:2020jkz}.}.

\section{Beyond the dilute approximation}

In the previous section, we derived the entanglement entropy
under the dilute approximation
\begin{equation}
\begin{split}
\langle b_a^i(\v{q})b_b^j(-\v{p}) \rangle \equiv
\mathcal{N} \int D \rho\, \,  e^{-\int_{\v{k}}\frac{1}{2\mu^2(\v{k})}\rho_c(\v{k})\rho^*_c(\v{k})}
 b_a^i(\v{q})b_b^j(-\v{p})
\approx
  \delta^2(\v{q}-\v{p})(2\pi)^2g^2\mu^2\delta^{ab}\frac{\v{p}^i\v{p}^j}{p^4}   \,.
\end{split}
\end{equation}

As we pointed out before, this approximation neglects any saturation corrections in the wave function
and as a result only the longitudinally polarized gluons contribute
to the entanglement entropy. Here, our goal is to take into account the saturation corrections.

In general the solution of the static Yang-Mills equation for the Weizsacker-Williams field is given by
\begin{equation}
    b^a_i(\v{x}) = \frac{1}{ig N_c} \Tr \left[T^a  U^+(\v{x}) \partial_i U(\v{x}) \right] = \frac{2}{ig} \Tr \left[t^a  V^+(\v{x}) \partial_i V(\v{x}) \right]\,.
\end{equation}
Given that the color charge distribution must be globally color invariant, the field correlator must  have the form  $ \langle b_a^i(\v{x}) b_b^j(\v{y}) \rangle \propto \delta^{ab}$.   Summing with respect to the colors we have
\begin{align}
\langle b^{a}_{i} (\v{x}) b^{a}_{j} (\v{y}) \rangle &=
 - \frac{4}{g^2}
 \langle \Tr \left[t^a  V^+(\v{x}) \partial_i V(\v{x}) \right]
 \Tr \left[t^a  V^+(\v{y}) \partial_j V(\v{y}) \right] \rangle
 \notag \\
 &=  - \frac{2}{g^2}
  \langle \Tr \left[  V^+(\v{x}) \partial_i V(\v{x}) V^+(\v{y}) \partial_j V(\v{y}) \right] \rangle
  \notag \\
  &= \frac{(2\pi)^3}{2} xG_{ij}^{\text WW}(x, \v{x} - \v{y})\,.
\end{align}
Here $xG^{ij}_{\text WW}(x, \v{x} - \v{y})$ is the Weizs\"acker-Williams gluon distribution  in the coordinate space.
Performing Fourier transformation and taking into account the translational invariance of the Weizs\"acker-Williams gluon distribution function, we arrive at
\begin{equation}
 \langle b^{a}_{i} (\v{k}) b^{a}_{j} (\v{q}) \rangle =
 \frac{ (2\pi)^{5} } { 2  }
  \delta^{(2)}(\v{k}+\v{q})
\int d^2r e^{-i \v{r}\cdot \v{k}} xG_{ij}^{\text WW}(x, \v{r})
 =
 \frac{ (2\pi)^{5} } { 2  }
 \frac{   \delta^{(2)}(\v{k}+\v{q}) } { S_\perp}
 xG^{ij}_{WW}(x,k)\,.
\end{equation}
The factor of the transverse area  $S_\perp$ in the denominator originates from the  commonly accepted definition of $xG^{ij}_{\text WW}(x,k)$:
\begin{equation}
    xG^{ij}_{\text WW}(x,k)  =
    \int  d^2x d^2y e^{-i (x-y) k}
    xG^{ij}_{\text WW}(x, \v{x} - \v{y})
     = S_\perp \int d^2r e^{-i \v{r}\cdot \v{k}} \, xG_{ij}^{\text WW}(x, \v{r})\,.
\end{equation}

Thus
\begin{equation}
 \langle b^{a}_{i} (\v{k}) b^{b}_{j} (\v{q}) \rangle =
 \frac{ (2\pi)^{5} } { 2 (N_c^2-1) S_\perp}   \delta^{(2)}(\v{k}+\v{q})  \delta^{ab} \, xG^{ij}_{\text {WW} }(x,k)
\end{equation}
The tensor $xG^{ij}_{\text{ WW} }(x,k)$ is conventionally split into two independent components:
\begin{align}
    xG^{ij}_{\text WW}(x,k) =
    \frac{1}{2} \delta_{ij} xG^{(1)} (x,k)
    - \frac{1}{2}  \left( \delta_{ij} - 2 \frac{k_i k_j} {k^2} \right)  xh^{(1)} (x,k)
\end{align}
where $xh^{(1)}$ is the linearly polarized gluon distribution.
Thus in general the Weizs\"acker-Williams field correlator contains both longitudinal and transverse components, with the transverse component proportional to $xG^{(1)}-xh^{(1)}$.

In the MV model both components can be computed semi-analitically \cite{Dominguez:2011br} to yield
\begin{align}
    \label{Eq:xh}
xh^{(1)}(x,q_{\perp})&=\frac{S_{\perp}}{2\pi^3\alpha_s}\frac{N_c^2-1}{N_c}\int_0^{\infty}dr_{\perp}\frac{r_{\perp}J_2(q_{\perp} r_{\perp})}{r_{\perp}^2\ln(\frac{1}{r^2_{\perp}\Lambda^2})}\left[1-e^{-\frac{1}{4}r^2_{\perp}Q_s^2\ln(\frac{1}{r^2_{\perp}\Lambda^2})}\right]\,,\\
\label{Eq:xG}
xG^{(1)}(x,q_{\perp})&=\frac{S_{\perp}}{2\pi^3\alpha_s}\frac{N_c^2-1}{N_c}\int_0^{\infty}dr_{\perp}\frac{r_{\perp}J_0(q_{\perp} r_{\perp})}{r_{\perp}^2}\left[1-e^{-\frac{1}{4}r^2_{\perp}Q_s^2\ln(\frac{1}{r^2_{\perp}\Lambda^2})}\right]\,,
\end{align}
where  $\Lambda$ is a non-perturbative IR scale, and the saturation momentum $Q_s$ is given by
\begin{align}
Q_s^2 = N_c \alpha_s g^2 \mu^2\,.
\end{align}
In this paper, we use the expression in the MV model for the Weizs\"acker-Williams gluon
field distributions; our results, however, can be straightforwardly extended to account
for the small-$x$ evolution~\cite{Dumitru:2015gaa}.

Now we are ready to revise the derivation of the reduced matrix to account for the
saturation corrections. We go back to the integration over the valence degrees of freedom
in Eq.~\eqref{Eq:Gaussian}:
\begin{align}
    \left\langle
        e^{- i\int_{\v{k}} b_a^i(\v{k}) (\phi^{i}_{1,a} (- \v{k}) - \phi^{i}_{2,a} (- \v{k}))   }
    \right\rangle
&=
1   + \sum_{q=1}^\infty \frac{1}{q!} \left\langle   \left[-i\int_{\v{k}} b_a^i(\v{k}) (\phi^{i}_{1,a} (- \v{k}) - \phi^{i}_{2,a} (- \v{k}))
\right]^q
\right\rangle \notag
\end{align}

The right hand side contains all higher order correlators of the Weizs\"acker-Williams field. We will however invoke a simple minded mean field Gaussian approximation in which all higher correlators factorize into products of the two point function. In this approximation we have
\begin{equation}
\left\langle
        e^{- i\int_{\v{k}} b_a^i(\v{k}) (\phi^{i}_{1,a} (- \v{k}) - \phi^{i}_{2,a} (- \v{k}))   }
    \right\rangle_{\text{MV}}=
e^{-\int_{\v{k}}\frac{1}{2}
\tilde M^{ab}_{ij}(\v{k})
  (\phi_{b1j}(\v{k}) - \phi_{b2j}(\v{k})) (\phi_{a1i}(-\v{k}) - \phi_{a2i}(-\v{k}))
   }\,,
\end{equation}
with
\begin{equation}
    \label{Eq:Mtilde}
    \tilde M^{ab}_{ij}(\v{k}) 
    =
    \frac{ (2\pi)^{3} \delta^{ab}} { 2(N_c^2-1)S_\perp  }    xG^{ij}_{WW}(x,k)\,.
\end{equation}
This approximation albeit simple, allows us to incorporate the main saturation effects in the CGC density matrix.
While deriving Eq.~\eqref{Eq:Mtilde} we took into account that $\delta^{(2)}(\v{k}=0) = S_\perp/(2\pi)^2$.

In the limit $k\gg Q_s$, we recover the dilute approximation of the previous section, i.e. $ \tilde M^{ab}_{ij}(k)  \to M^{ab}_{ij}(k)$. To show this explicitly consider the eigenvalues of   $xG^{ij}_{WW}(x,k)$. Those are     $(xG^{(1)} \pm xh^{(1)})/2$. At large momentum one can consider small $r$ in the integrals \eqref{Eq:xh} and \eqref{Eq:xG}. Expanding the exponentials, we obtain
\begin{equation}
\begin{split}
xh^{(1)}(x,q_{\perp})&\approx\frac{S_{\perp}}{2\pi^3\alpha_s}\frac{N_c^2-1}{N_c}\frac{Q_s^2}{4}\int_0^{\infty}dr_{\perp}r_{\perp}J_2(q_{\perp} r_{\perp}) \approx \frac{S_{\perp}}{4 \pi^3\alpha_s}\frac{N_c^2-1}{N_c}\frac{Q_s^2}{q^2}\, , \\
xG^{(1)}(x,q_{\perp})&\approx\frac{S_{\perp}}{2\pi^3\alpha_s}\frac{N_c^2-1}{N_c}\frac{Q_s^2}{4}\int_0^{\infty}dr_{\perp}r_{\perp}J_0(q_{\perp} r_{\perp})\ln(\frac{1}{r^2_{\perp}\Lambda^2}) \approx
\frac{S_{\perp}}{4 \pi^3\alpha_s}\frac{N_c^2-1}{N_c}\frac{Q_s^2}{q^2}
\,.
\end{split}
\end{equation}
This reproduces the limits discussed in Ref.~\cite{Metz:2011wb}.
We thus obtain one zero eigenvalue and the other one given by
\begin{align}
\lim_{q\gg Q_s}\frac{xG^{(1)}+xh^{(1)}}{2}=\frac{S_{\perp}}{4\pi^3\alpha_s}\frac{N_c^2-1}{N_c}\frac{Q_s^2}{q^2_{\perp}}\,.
\end{align}
The nontrivial eigenvalue of $\tilde M$ becomes
\begin{align}
\frac{S_{\perp}}{4\pi^3\alpha_s}\frac{1}{N_c}\frac{Q_s^2}{q^2_{\perp}} \times \frac{ (2\pi)^{3} \delta^{ab}} { 2S_\perp  } =\frac{ g^2\mu^2\delta^{ab}}{q_{\perp}^2}\,.
\end{align}
This indeed reduces to the expression in dilute limit used in the previous section.

Repeating the derivation of the previous section separately for each eigenvalue of  of  the matrix $\tilde M$, we obtain
\begin{align}\label{ent1}
    S^E&=
    \frac{N_c^2-1}{2} \sum_{\nu=\pm}
    \int_\v{k}
    \left[ \ln{\tilde M_\nu}(\v{k})+\sqrt{1+4\tilde  M_\nu(\v{k})}\ln(1+\frac{1}{2 \tilde  M_\nu(\v{k})}+\frac{\sqrt{1+4\tilde M_\nu(\v{k})}}{2 \tilde M_\nu(\v{k})}
    )\right]
\end{align}
where
\begin{align}\label{mpm}
    \tilde  M_{\pm} = \frac{ (2\pi)^{3} } { 2 S_\perp (N_c^2-1) }  \frac{xG^{(1)} \pm xh^{(1)}}{2}\,.
\end{align}

Comparing to Eq.~\eqref{Eq:ent} there are two major differences. First, Eq.~\eqref{ent1} contains a nontrivial contribution of the transverse mode, and second the small momentum behavior of the larger eigenvalue is very different from that of $M_+$. We will return to the discussion of this point later on.

\section{Diagonalization of the reduced density matrix}

\subsection{The Boltzmann density matrix}
Interestingly, the result of the previous section can be rewritten in a very suggestive form~\footnote{A similar result was obtained in a general Gaussian density matrix~\cite{Berges:2017hne}.}
\begin{align}\label{Eq:bol}
    S_E = (N_c^2-1) S_\perp \sum_{\nu=\pm}
    \int \frac{d^2k}{(2\pi)^2} \Bigg[
    (1+f_\nu) \ln (1+f_\nu) - f_\nu \ln f_\nu
    \Bigg]\,.
\end{align}
Here we  defined the distribution functions
\begin{align}
    f_\pm (k)= \frac{1}{\exp(\beta \omega_\pm(k)) -1 }
\end{align}
with
\begin{align}
    \label{Eq:betaomega}
\beta \omega_\pm(\v{k})  = 2 \ln(\frac{1}{2\sqrt{\tilde M_\pm(\v{k})}}+\sqrt{1+\frac{1}{4 \tilde  M_\pm(\v{k})}})\,.
\end{align}

This suggests that in a basis in which the density matrix is diagonal, it must have the  Boltzmann form,
\begin{align}
   \hat \rho= N  e^{- \beta \omega_+ \hat n_+  } e^{- \beta \omega_- \hat n_-  },
\end{align}
where $\hat n_\pm$ is the corresponding number density of the quasi particles. We will refer to this basis as the quasiparticle basis.

Our purpose in this section is to find the quasiparticle basis explicitly. Before turning to this problem, we ask what is the dispersion relation of the quasiparticles provided our interpretation of Eq.~\eqref{Eq:bol} is correct.

Let us first examine the dilute case of Section II.
For the only nontrivial polarization  $M_+= g^2 \mu^2 /k^2$ and at small momenta $k\ll g^2\mu^2$  Eq.~\eqref{Eq:betaomega} gives $\beta \omega_+ \approx k/(g\mu)$. Interestingly this looks like a dispersion relation of a massless particle. Although only the product of the frequency and the inverse temperature is determined by Eq.~\eqref{Eq:betaomega}, assuming the velocity of quasiparticles is the speed of light,  the inverse temperature is  $\beta=(g\mu)^{-1}$.  At large momentum $\beta \omega_+ \approx \ln ( k^2/g^2\mu^2)$ or $f_+ \approx  g^2\mu^2/k^2$. This perturbative-like behavior is then interpreted as a logarithmic dispersion relation for quasiparticles at high momenta.

One interesting point to note is that the transition between the ``low momentum'' and ``high momentum'' regimes in the present context  is given by the scale $g^2\mu^2$ which is {\it parametrically larger} than the saturation momentum $Q_s$, i.e. $g^2\mu^2=Q_s^2/\alpha_sN_c$. Physically this is easy to understand. While saturation effects become important at the momentum scale at which the gluon occupation number is large, of order $1/\alpha_s$, the perturbative regime, i.e. momenta for which the eigenvalue $M_+$ becomes small requires the occupation number to be smaller than unity. Thus there is a whole range of momenta, at which the gluon occupation number is greater than one, but saturation effects are still unimportant. We will refer to this range of momenta, $Q^2_s<k^2< Q_s^2/\alpha_sN_c$ as ``semi hard''. At small 'tHooft coupling $\alpha_sN_c$ the semi hard region is parametrically large.


The saturation corrections discussed in the previous section have a strong effect on the ``dispersion relation'' at low momenta. Beyond the dilute limit
$xG^{(1)}$ dominates over $xh^{(1)}$ at $k<Q_s$, so that both eigenvalues $\tilde M_\pm$ behave as $xG^{(1)} \propto \ln Q_s/k$ in the infrared. This is a well known effect where the infrared $1/k^2$ behavior of the TMD which formally leads to a power like infrared divergence in the produced particle spectrum is tamed by the saturation corrections and becomes logarithmic. This correction results in the logarithmic ``dispersion relation''   $\beta \omega_\pm \propto
\sqrt{\ln Q_s/k} $.
This modification however is only important at very low momenta $k^2<Q_s^2$. In the semi hard regime $Q^2_s<k^2< Q_s^2/\alpha_sN_c$  the quasiparticles corresponding to the eigenvalue $M_+$ still
behave approximately as massless bosons.  For these momenta the eigenvalue $M_+$ is large.  For large $M_+$ we can expand logarithm in
Eq.~\eqref{Eq:betaomega}; which leads to $\beta \omega_+ \approx  1/\sqrt{ \tilde M_+ }$. Then taking  the high momentum approximation, that is $ \tilde M_+ \propto Q_s^2/\alpha_sk^2$ we obtain  $\beta \omega_+ \propto \sqrt{\alpha_s}k/Q_s$, i.e. thermal spectrum for massless particles. The temperature is of the order $T\sim Q_s/\sqrt{\alpha_sN_c}$ which is parametrically larger than the saturation momentum. The reason, as we explained above is that the gluon occupation number is of order unity for momenta which are much higher than $Q_s$, and it is the value of the momentum of the highest occupied state that determines the effective temperature.

\subsection{Construction of the eigenvalue problem}
We now perform the explicit diagonalization of the reduced density matrix.

First we note  that the density matrix has a product form in the momentum space due to the factorization of the momentum modes.
Using Eq.~\eqref{Eq:rhoME} we can write down  the density matrix operator as
\begin{align}
    \hat\rho=\prod_{\v{k}} \hat{\rho}_\nu(\v{k})
    =&\prod_{\nu=\pm}\prod_\v{k}{\cal N}\int D\phi(\v{k}) D\Phi(\v{k}) \notag \\&\times e^{-\frac{1}{4}(\phi(\v{k}) \phi(-\v{k}) +\Phi(\v{k})\Phi(-\v{k}))} e^{-\frac{1}{2}\tilde M\phi(\v{k}) \phi(-\v{k}) -\frac{1}{2} \tilde M\Phi(\v{k}) \Phi(-\v{k}) }e^{\tilde M \phi(\v{k}) \Phi(-\v{k}) }
\notag
    \\
    & \times
    |\phi(\v{k})\rangle \langle \Phi(\v{k})| \,,\label{D1}
\end{align}
where $\tilde M$ is an eigenvalue of $\tilde M^{ij}$
\begin{align}
    \tilde M=\tilde M_+ \,\, \text{or}\,\,  \tilde M_-\,.
\end{align}
Equation~\eqref{D1} should also contain a product over the index $\nu=\pm$, which we do not indicate explicitly.

Since $\hat\rho$ is a product over momentum (and polarization) we will consider only a single momentum (and polarization) mode for the sake of simplifying the notations.

The Boltzmann form of the entanglement entropy suggests that all eigenvalues of the density matrix are given by integer powers of the same number.
Our goal is to find these eigenvalues explicitly by solving the eigenvalue equation
 \begin{align}
      \hat{\rho}|\Psi_i\rangle=\lambda_i |\Psi_i\rangle\,.
 \end{align}
We write the eigenstate in the field basis as
\begin{align}
    |\Psi_i(\v{k})\rangle   = \int d\phi(\v{k})  f_i(\phi(\v{k})) | \phi(\v{k}) \rangle .
\end{align}
The eigenvalue equation for the wave function $f_i$ becomes
\begin{align}\label{eigene}
   \frac{1}{\sqrt{2\pi}} \int d \Phi   e^{- \frac{(1+2M)}{4}(\phi(-\v{k})\phi(\v{k})+\Phi(-\v{k})\Phi(\v{k}))}e^{M \phi(-\v{k}) \Phi(\v{k}) } f_i(\Phi)=\lambda_i f_i(\phi)\,.
\end{align}
The form of the integrand suggests to look for the ground state in the form of the Gaussian
\begin{align}\label{Eq:vac}
f_0(\phi(\v{k})) = N \exp \left[ - \alpha \phi(-\v{k}) \phi(\v{k})  \right]\,.
\end{align}
Once we find the constant $\alpha$ by solving Eq.~\eqref{eigene}, we will generate exciting states by acting with the
the creation operators, which is defined analogously to the quantum oscillator problem
\begin{align}
    \label{Eq:Cr}
    &c(\v{k})=\frac{1}{\sqrt{2}}\left(\sqrt{\alpha}\phi(\v{k})+\frac{(2\pi)^2}{\sqrt{\alpha}}\frac{\delta}{\delta \phi(-\v{k})}\right)\,,\\
&c^{\dagger}(\v{k})=\frac{1}{\sqrt{2}}\left(
    \sqrt{\alpha} \phi(-\v{k}) - \frac{(2\pi)^2}{\sqrt{\alpha}} \frac{\delta}{\delta \phi(\v{k})}\right)\,.
\end{align}
The operators satisfy
\begin{align}
    \left[ c(\v{p}) , c^{\dagger}(\v{k})\right]&=(2\pi)^2\delta^{(2)}(\v{p}-\v{k})\,,\\
    c(\v{k})f_0(\phi(\v{k}))&=0\,.
\end{align}
It can be checked straightforwardly that once the appropriate $\alpha$ is found, the states obtained by repeated action of $c^\dagger(k)$ on the Gaussian state Eq.~\eqref{Eq:vac} \begin{align}
    f_n(\phi(\v{k}))=\left(\frac{\alpha}{2}\right)^{\frac{n}{2}}\phi(-\v{k})^n \exp
    \left[-\alpha \phi(\v{k})\phi(-\v{k})\right]
\end{align}
are in fact eigenstates of the density matrix.


Substituting the Gaussian function Eq.~\eqref{Eq:vac} into Eq.~\eqref{eigene} after a little algebra we find
\begin{align}
    4 \alpha^\pm&=\sqrt{1+4 \tilde M_\pm} \\
\lambda^\pm_0 &=\frac{\sqrt{2}}{\sqrt{1+2\tilde  M_\pm +4\alpha^\pm}}=\frac{2}{1+\sqrt{1+4\tilde M_\pm}}\,.
\end{align}



Proceeding similarly for the excited states we find
\begin{align}
    \lambda^\pm_n  =\left[\frac{2\tilde M_\pm}{\left(1+\sqrt{1+4\tilde M_\pm}\right)^2}\right]^n\lambda_0^\pm\,.
\end{align}

This results confirms our earlier expectation that the density matrix has the Boltzmann form. In terms of the operators $c$ and $c^\dagger$ it can be written as
\begin{equation}
\hat\rho(\v{k})=N\left[\frac{2\tilde M_\pm}{\left(1+\sqrt{1+4\tilde M_\pm}\right)^2}\right]^{c^\dagger c}=Ne^{-\beta\omega(\v{k})c^\dagger(\v{k})c(\v{k})}
\end{equation}
with
\begin{equation}
\beta\omega=\ln \left[\frac{\left(1+\sqrt{1+4\tilde M_\pm}\right)^2}{2\tilde M_\pm}\right]
\end{equation}
and $N$ - the appropriate normalization factor.
This coincides with Eq.~\eqref{Eq:betaomega}.

\subsection{The Bogoliubov transformation}

We have thus explicitly established that in the basis of the quasiparticles defined by the creation and annihilation operators
$c^{\dagger}(k)$ and $c(k)$ ,  the density matrix is diagonal and has a Boltzmann form. It is easy to show that the quasiparticle basis is related to the perturbative gluon Fock space basis by a simple Bogolyubov transformation:
\begin{align}
    c_\pm(k) &= \cosh (B_\pm) \, a_\pm(k) +  \sinh (B_\pm) \, a_\pm^\dagger(-k)\,, \\
    c_\pm^\dagger(k) &=  \cosh (B_\pm) \,  a_\pm^\dagger (k) + \sinh (B_\pm) \, ,a_\pm(-k)
\end{align}
where $B_\pm = \ln 2 \sqrt{\alpha_\pm}  =  \frac 1 4\ln \left( 1+4\tilde M_\pm \right)$.\\
Here we have restored the polarization indices, and have defined $a_+(k)\equiv\hat k_ia_i(k)$ and $a_-\equiv \epsilon_{ij}\hat k^ia_j(k)$ with $\hat k$ - a unit vector in the direction of the transverse momentum $k$.

How ``far'' is the quasiparticle Fock space removed from the perturbative gluon Fock space? The answer clearly depends on the value of transverse momentum. Let us consider the two simple limiting cases:
\begin{itemize}
 \item[$k\gg Q_s$:] here we have $\tilde M_-\propto k^{-4}$ while $\tilde M_+ \propto k^{-2}$. For either polarization  $B_{\pm} \approx 0$ so that the quasiparticle basis practically coincides with the perturbative gluon basis $ c_\pm(k) \approx \, a_\pm(k)  $. This is natural since for large momenta the occupation number of gluons vanishes, and the density matrix in the first approximation is just given by the perturbative vacuum.

  \item[$k\ll Q_s$:]  The situation is quite different in this limit. Here $\tilde M_\pm$ is large, and $\sinh (B_{\pm}) \simeq \cosh (B_{\pm}) \simeq e^{B_{\pm}}$; the transformation in this case corresponds to maximal mixing. Interestingly, this maximal mixing regime only requires that $\tilde M\gg 1$. Thus it is not only valid for very small momenta, but also for a considerably large range of ``semi hard'' momenta, $k<2\pi Q_s$. This is the same momenta for which the dispersion relation of the quasiparticles is approximately linear.
\end{itemize}

\section{Discussion}
In this paper we have extended the approach to CGC density matrix pioneered in \cite{Kovner:2015hga} by including saturation corrections in a  mean-field approximation. The effect of these corrections is two prong. First, they result in excitation of the transverse gluon mode, so that both gluon polarizations now contribute to entanglement entropy. Second, the infrared behavior of the Weizsacker-Williams field propagator is softened.

We have also pointed out that the reduced soft gluon density matrix can be explicitly diagonalized by a Bogoliubov transformation. In the quasi particle basis, it has a Boltzman form, i.e. for a given transverse momentum and polarization, its eigenvalues are powers of one number. If interpreted as a thermal density matrix, this determines the product of the quasiparticle energy and the inverse temperature. We found that for semi hard momenta $Q_s^2<k^2<Q_s^2/\alpha_sN_c$, the dispersion relation of the quasiparticles is approximately linear with momentum, while at very low momenta $k^2<Q_s^2$ the saturation effects lead logarithmic dispersion relation. The saturation then induces an effective mass for the quasiparticles, albeit this mass is not fixed, but rather runs with the momentum into the infrared. 

We also noted that the effective temperature for the quasiparticle system is not given by the saturation momentum $Q_s$, but rather by a parametrically greater scale $T\sim Q_s/\sqrt{\alpha_sN_c}$. The physical reason is that the temperature is determined by the momentum of those levels for which the occupation number is of order unity, rather than much larger than unity.

The two distinct scales that arise in the physics of saturated wave function is reminiscent of the two scales present at high temperature in weakly interacting quark-gluon plasma. The softest scale -- the saturation momentum is closely analogous to the so called ``magnetic mass''. Both  arise due to self interaction of very soft modes, and  both are parametrically $m_{\rm soft}\propto \alpha_s \Lambda_{\rm hard}$, where in the case of plasma $\Lambda_{\rm hard}=T$, while in CGC $\Lambda_{\rm hard}=\mu$. The semi hard scale, parametrically $m_{\rm semi\ hard}\propto g\Lambda_{\rm hard}$ is identified with the ``electric mass'' in plasma and the effective temperature in CGC. This scale arises in both cases via the interaction of semi hard modes with the hard ones. In the case of plasma the relevant mechanism is ``hard thermal loops'', while in the case of CGC it is the eikonal emissions by the valence charges that populate the CGC wave function in the semi hard region. Amusingly, in both case only ``electric'' modes are affected by this scale, in the sense that in the plasma the electric mass produces finite correlation length only for the chromoelectric field, while in the CGC wave function only the electric (longitudinal in the two dimensional sense) modes are populated in the semi hard region.

Note that the value of the effective temperature in this paper is very different from the one obtained in  \cite{Kovner:2015hga}. The reason is that in  \cite{Kovner:2015hga} the inverse temperature was defined as the derivative of the entropy with respect to transverse energy. The total transverse energy is dominated by the UV modes despite the low occupancy of each energy level. Thus the temperature calculated this way in  \cite{Kovner:2015hga} was dominated by the contribution of the UV modes and came out proportional to the UV cutoff.  Conversely when considering properties of the density matrix of {\it produced gluons}, which is dominated by momenta of order $Q_s$, the temperature in  \cite{Kovner:2015hga} came out to be of order $Q_s$. In the present paper however we discussed the effective temperature of those modes which have an approximate Boltzmann distribution of massless bosons. These turned out to be semi-hard modes, and the temperature accordingly turned out to be tied to the appropriate semi hard scale.

We conclude by noting that the concept of gluon quasi particles inside a hadron is a very interesting concept. It is especially worth stressing, that while at high momenta the quasi particles coincide with perturbative gluons, at semi-hard momenta the quasi particle operators are very different from the perturbative gluon creation and annihilation operators as is clear from our discussion in the previous section. Thus the semi hard momentum region arises here as a well defined transition region between the perturbative hard and a genuinely nonperturbative soft ($k<Q_s$) regimes. One is reminded of the Landau liquid theory, where quasiparticles indeed arise via possibly strong dressing of original particles while still retaining the same quantum numbers and a particle identity.

It would be extremely interesting to understand how to probe experimentally the properties of such gluon quasi particles. This is of course a very difficult question, since the quasi particles discussed here exist inside the wave function of a hadron, while any final state is described (modulo hadronization corrections) in terms of original perturbative gluons. Nevertheless it seems to us that this problem is well worth thinking about.

\begin{acknowledgments}
We acknowledge usefull conversation with M.~Li, M.~Lublinsky, and E. Levin.
We are especially thankfull to R. Venugopalan for illuminating discussions.

A.K. is supported by the NSF Nuclear Theory grant 1913890.  This material is based upon work supported
by the U.S. Department of Energy, Office of Science,
Office of Nuclear Physics through the Contract No. DE-SC0020081 (H.D. and V.S.).
\end{acknowledgments}

\end{document}